\documentstyle[12pt]{article}

\newcommand {\e} {\mbox{\rm e}}

\newcounter{eq}
\newcounter{sc}

\def\overleftrightarrow#1{\vbox{\ialign{##\crcr
 $\leftrightarrow$\crcr\noalign{\kern-1pt\nointerlineskip}
 $\hfil\displaystyle{#1}\hfil$\crcr}}}

\newlength{\minitwocolumn}
\begin{document}

\begin{flushright}
DPUR/TH/55\\
June, 2017\\
\end{flushright}
\vspace{20pt}

\pagestyle{empty}
\baselineskip15pt

\begin{center}
{\large\bf Quantum Aspects of Nonlocal Approach to the Cosmological Constant Problem
\vskip 1mm }

\vspace{10mm}
Ichiro Oda \footnote{E-mail address:\ ioda@phys.u-ryukyu.ac.jp
}

\vspace{3mm}
           Department of Physics, Faculty of Science, University of the 
           Ryukyus,\\
           Nishihara, Okinawa 903-0213, Japan.\\

\end{center}


\vspace{3mm}
\begin{abstract}
We have recently presented a manifestly local and general coordinate invariant formulation of a nonlocal approach
to the cosmological constant problem which has been proposed by Carroll and Remmen.
In this article, based on our formulation, we investigate some quantum aspects of this approach
to the cosmological constant problem.
In particular, we explicitly evaluate quantum effects to the effective cosmological constant from matter fields 
at one-loop level, and show that the effective cosmological constant receives radiative corrections and is not stable 
against quantum corrections so that it must be fine-tuned at every stage in the loop expansion. 
Next, we will propose a new geometrical model of a nonlocal approach to the cosmological constant problem, 
and show explicitly that in this new model the effective cosmological constant is indeed stable against radiative 
corrections. 
\end{abstract}

\newpage
\pagestyle{plain}
\pagenumbering{arabic}


\rm
\section{Introduction}

The discovery of accelerated expansion of the universe by two astrophysical groups in 1998 \cite{Riess, Perlmutter}
has pushed the cosmological constant problem to the forefront of research in modern theoretical physics.  
This discovery was achieved by measuring the apparent luminosity of distant supernovae. Their dimness implies that
the accelerated expansion of the universe has recently (according to the cosmological standard) started
and it turns out to be compatible with a very small, positive cosmological constant and be inconsistent with
the vanishing cosmological constant.

For a long time before its discovery, most of theoretical physicists have unsuccessfully attempted to
construct a theory which explains why the cosmological constant, or equivalently, the vacuum energy density,
vanishes precisely by some still-unknown dynamical mechanism or symmetry. However, after the discovery,
this situation has completely changed and we are now facing a more complicated problem, that is, we have to
account for not only why the cosmological constant in the universe is not exactly zero but very tiny,
but also why the present size of the cosmological constant is of the same order of magnitude as the
energy density of ordinary matters.

Trying to find a solution to the cosmological constant problem, there is an important no-go theorem by Weinberg
based on classical physics \cite{Weinberg}, which states that no local field equations including classical gravity can have 
a flat Minkowski space-time as classical solutions for generic values of parameters, or to put differently,
any solution to the cosmological constant problem cannot be obtained without fine-tuning within the local field 
theories including gravity. Thus, to bypass this no-go theorem, it seems to be unavoidable to turn our attention
to some kinds of nonlocal field theories.  Recently, there has appeared such a nonlocal approach to 
the cosmological constant problem by Carroll and Remmen \cite{Carroll} where a nonlocal constraint 
equation such that the total action is vanishing was imposed in order to evade the Weinberg theorem.
However, this nonlocal approach is somewhat ad hoc and phenomenological in that the nonlocal constraint
equation is derived from the variation with respect to a mere constant number.

More recently, to remedy this defect, we have constructed a manifestly local formulation to the nonlocal
approach by Carroll and Remmen \cite{Oda0}. In our local formulation, the same constraint equation as that of
Carroll and Remmen can be obtained in terms of a local Lagrange multiplier field by adding an additional
topological term in the action.

In both the formulations, however, only the classical analysis has been done, so it is the main
purpose in this article that we wish to shed light on quatum aspects of this nonlocal approach.
In particular, we would like to check whether or not the effective cosmological constant obtained in our previous work
\cite{Oda0}, which is truly a cosmological constant controlling the Friedmann equation etc., is stable
under quantum effects of matter fields, and in addition to it, clarify the physical meaning of the effective cosmological 
constant at the quantum level. 

Before doing that, it is worthwhile to comment on some quantum aspects of the cosmological constant
problem which are relevant to the study at hand. Let us recall that fundamentally, the cosmological 
constant problem amounts to a clash between particle physics which sources the vacuum energy
density through quantum effects and gravity which responds to it classically. To describe this
physical situation more accurately, we might need a theory of quantum gravity, but it should be
noticed that quantum gravity would be in general needed to describe physics around the Planck
mass scale $10^{19} GeV$ or curvature radii smaller than the Planck length $10^{-33} cm$.
Hence, in considering the cosmological constant problem, it is not always necessary to take into
consideration such extreme regimes, and the cosmological constant problem already arises
well inside the regime of validity of both classical gravity and quatum field theory, in other words,
inside the semiclassical regime. Accordingly, in this article, while matter fields are treated
quantum mechanically, gravity is treated classically.  An important problem associated with
the cosmological constant problem is that quantum effects coming from fluctuations of matter fields
could render the size of the cosmological constant infinite in the limit of the infinite cutoff since 
the cosmological constant would receive a contribution of order $\Lambda_{cutoff}^4$ in the absence of
fine-tuning and some symmetry when we use the ultraviolet cutoff $\Lambda_{cutoff}$ as the regularization procedure.   
Then, the cosmological constant problem becomes a hierarchy
problem of explaining the large discrepancy between the observed low value of the vacuum energy
density and the theoretical large value of it. In this article, by the explicit calculation, we wish to verify
how the effective cosmological constant receives corrections when quantum effects of matter fields 
are added.    

This paper is organised as follows: In Section 2, we review a manifestly local and generally coordinate 
invariant formulation which reduces to the action by Carroll and Remmen up to the higher-derivative curvature terms. 
In Section 3, we derive all the equations of motion and show that the space-time average of the gravitational 
field equations produces very similar equations to those obtained Carroll et al. In Section 4, we evaluate the effective 
cosmological constant when quantum effects of matter fields are added, and see that we need the fine-tuning of the
cosmological constant at the one-loop level, which means the instability against radiative corrections. In Section 5, 
we propose a new geometrical model and show that this new model is free of the radiative instability problem
and provide us with a new perspective to the cosmological constant problem. The final section is devoted to discussions.

\section{Review of manifestly local formulation}

In this section we wish to not only review a manifestly local and generally coordinate invariant formulation \cite{Oda0} for a nonlocal approach 
to the cosmological constant problem by Carroll and Remmen \cite{Carroll}, but also slightly generalize it by including higher-derivative curvature 
terms which are needed for renormalization.  

Since we would like to preserve the main feature of the original work
by Carroll et al.  \cite{Carroll}, the sector which we add should not gravitate directly. The unique and nontrivial possibility would be to add 
a topological term to the Carroll and Remmen action \cite{Oda0}. \footnote{The idea to follow has been already suggested in the gauge invariant 
formulation of unimodular gravity \cite{Einstein, Weinberg} by Henneaux and Teitelboim \cite{Henneaux}. This idea has been also used to construct 
topological induced gravity \cite{Oda1, Oda2, Oda3} and a manifestly local, diffeomorphism invariant and locally 
Poincar\'e invariant formulation \cite{Kaloper2} of vacuum energy sequestering \cite{Kaloper1, Kaloper3}. }

A manifestly local and generally coordinate invariant action for the Carroll-Remmen approach \cite{Carroll} has
been already proposed in \cite{Oda0} \footnote{We follow notation and conventions of the textbook by Misner et al \cite{MTW}.} 
\begin{eqnarray}
S = S_{CR} + S_{Top},
\label{Our Action}
\end{eqnarray}
where each action in the right hand side (RHS) is defined as follows: \footnote{The last boundary term is included
to obtain stationary action under variations that leave the field strength $F_{\mu\nu\rho\sigma}$ fixed on the boundary 
\cite{Duncan}.} 
\begin{eqnarray}
S_{CR}  &=& \int d^4 x \sqrt{- g} \ \eta(x) \Biggl[ \frac{1}{16 \pi G_b}  ( R - 2 \Lambda_b ) 
+ k_{1b} R^2 + k_{2b} R_{\mu\nu}^2 + k_{3b} R_{\mu\nu\rho\sigma}^2   
\nonumber\\
&+& {\cal{L}}_{mb} - \frac{1}{2} \cdot \frac{1}{4!} F_{\mu\nu\rho\sigma}^2 
+ \frac{1}{6} \nabla_\mu ( F^{\mu\nu\rho\sigma} A_{\nu\rho\sigma} )
\Biggr],
\label{CR Action}
\end{eqnarray}
and 
\begin{eqnarray}
S_{Top}  = \int d^4 x  \ \frac{1}{4!}  \ \mathring{\varepsilon}^{\mu\nu\rho\sigma} \eta(x) H_{\mu\nu\rho\sigma}.
\label{Top Action}
\end{eqnarray}

In the above, we have introduced various quantities: $g$ is the determinant of the metric tensor, $g = \det g_{\mu\nu}$, and $R$ denotes the
scalar curvature. $\eta(x)$ is a scalar field, which reduces to the constant Lagrange multiplier parameter of Carroll et al. after taking a classical
solution. $G_b$, $\Lambda_b$, and $ {\cal{L}}_{mb}$ are the bare Newton's constant, the bare cosmological constant and the bare Lagrangian density 
for generic matter fields, respectively, and $k_{ib} (i = 1, 2, 3)$ are bare coefficients in front of the higher-derivative curvature terms. 
Moreover, $F_{\mu\nu\rho\sigma}$ and $ H_{\mu\nu\rho\sigma}$ are respectively the field strengths for two 3-form gauge fields 
$A_{\mu\nu\rho}$ and $B_{\mu\nu\rho}$
\begin{eqnarray}
F_{\mu\nu\rho\sigma} = 4 \partial_{[\mu} A_{\nu\rho\sigma]},  \qquad 
H_{\mu\nu\rho\sigma} = 4 \partial_{[\mu} B_{\nu\rho\sigma]},
\label{4-forms}
\end{eqnarray}
where the square brackets denote antisymmetrization of enclosed indices. Finally, $\mathring{\varepsilon}^{\mu\nu\rho\sigma}$
and $\mathring{\varepsilon}_{\mu\nu\rho\sigma}$ are the Levi-Civita tensor density defined as
\begin{eqnarray}
\mathring{\varepsilon}^{0123} = + 1,  \qquad 
\mathring{\varepsilon}_{0123} = - 1,
\label{Levi}
\end{eqnarray}
and they are related to the totally antisymmetric tensors $\varepsilon^{\mu\nu\rho\sigma}$ and $\varepsilon_{\mu\nu\rho\sigma}$ via
relations
\begin{eqnarray}
\varepsilon^{\mu\nu\rho\sigma} = \frac{1}{\sqrt{-g}}  \mathring{\varepsilon}^{\mu\nu\rho\sigma},  \qquad 
\varepsilon_{\mu\nu\rho\sigma} = \sqrt{-g}  \mathring{\varepsilon}_{\mu\nu\rho\sigma}.
\label{Levi-tensor}
\end{eqnarray}
Also note that the Levi-Civita tensor density satisfies the following equations:
\begin{eqnarray}
\mathring{\varepsilon}^{\mu\nu\rho\sigma} \mathring{\varepsilon}_{\alpha\beta\rho\sigma} = - 2 ( \delta_\alpha^\mu \delta_\beta^\nu
-   \delta_\alpha^\nu \delta_\beta^\mu ),  \qquad 
\mathring{\varepsilon}^{\mu\nu\rho\sigma} \mathring{\varepsilon}_{\alpha\nu\rho\sigma} = - 3!  \delta_\alpha^\mu,  \qquad 
\mathring{\varepsilon}^{\mu\nu\rho\sigma} \mathring{\varepsilon}_{\mu\nu\rho\sigma} = - 4!.
\label{e-identity}
\end{eqnarray}

We will now show that our action becomes equivalent to that of Carroll et al. up to the higher-derivative terms.
For this purpose, let us take the variation with respect to the 3-form $B_{\mu\nu\rho}$, which gives us the equations 
for a scalar field $\eta(x)$
\begin{eqnarray}
\mathring{\varepsilon}^{\mu\nu\rho\sigma} \partial_\sigma \eta(x)  = 0,
\label{eta-eq}
\end{eqnarray}
from which we have a classical solution for $\eta(x)$
\begin{eqnarray}
\eta(x)  = \eta,
\label{eta-sol}
\end{eqnarray}
where $\eta$ is a certain constant. Substituting this solution into the starting action (\ref{Our Action}),
we arrive at the action proposed by Carroll et al. up to the higher-derivative terms to solve the cosmological constant 
problem \cite{Carroll}:
\begin{eqnarray}
S &=& \eta \int d^4 x \sqrt{- g} \  \Biggl[ \frac{1}{16 \pi G_b}  ( R - 2 \Lambda_b ) 
+ k_{1b} R^2 + k_{2b} R_{\mu\nu}^2 + k_{3b} R_{\mu\nu\rho\sigma}^2   
\nonumber\\
&+& {\cal{L}}_{mb} - \frac{1}{2} \cdot \frac{1}{4!} F_{\mu\nu\rho\sigma}^2 
+ \frac{1}{6} \nabla_\mu ( F^{\mu\nu\rho\sigma} A_{\nu\rho\sigma} )
\Biggr].
\label{Original CR Action}
\end{eqnarray}
Note that with Eq. (\ref{eta-sol}), the topological term $S_{Top}$ becomes a surface term, which is ignored in this derivation.

\section{Equations of motion}

Let us derive all the equations of motion from the action (\ref{Our Action}) which are completely local. 
Now the variation of the action (\ref{Our Action}) with respect to the scalar field $\eta(x)$ provides us with the vanishing
total Lagrangian density
\begin{eqnarray}
&{}& \sqrt{- g} \Biggl[ \frac{1}{16 \pi G_b}  ( R - 2 \Lambda_b ) 
+ k_{1b} R^2 + k_{2b} R_{\mu\nu}^2 + k_{3b} R_{\mu\nu\rho\sigma}^2  + {\cal{L}}_{mb}
\nonumber\\ 
&-& \frac{1}{2} \cdot \frac{1}{4!} F_{\mu\nu\rho\sigma}^2 
+ \frac{1}{6} \nabla_\mu ( F^{\mu\nu\rho\sigma} A_{\nu\rho\sigma} ) \Biggr] 
+ \frac{1}{4!}  \ \mathring{\varepsilon}^{\mu\nu\rho\sigma} H_{\mu\nu\rho\sigma} = 0.
\label{Zero-Lagr}
\end{eqnarray}
Next, we set $H_{\mu\nu\rho\sigma}$ to be
\begin{eqnarray}
H_{\mu\nu\rho\sigma}  = c(x) \varepsilon_{\mu\nu\rho\sigma} = c(x) \sqrt{-g} \mathring{\varepsilon}_{\mu\nu\rho\sigma},
\label{H}
\end{eqnarray}
where $c(x)$ is some scalar function. Then, using the last equation in Eq. (\ref{e-identity}), Eq.  (\ref{Zero-Lagr}) can be rewritten as
\begin{eqnarray}
&{}& \frac{1}{16 \pi G_b}  \left[  R - 2 \Lambda_b (x) \right] + k_{1b} R^2 + k_{2b} R_{\mu\nu}^2 + k_{3b} R_{\mu\nu\rho\sigma}^2
+ {\cal{L}}_{mb} 
\nonumber\\
&-& \frac{1}{2} \cdot \frac{1}{4!} F_{\mu\nu\rho\sigma}^2 + \frac{1}{6} \nabla_\mu ( F^{\mu\nu\rho\sigma} A_{\nu\rho\sigma} )
= 0,
\label{Zero-Lagr 2}
\end{eqnarray}
where we have defined 
\begin{eqnarray}
\Lambda_b (x) = \Lambda_b + 8 \pi G_b c(x).
\label{Lambda(x)}
\end{eqnarray}
The equations of motion for the 3-form $A_{\mu\nu\rho}$ take the form
\begin{eqnarray}
\nabla^\mu F_{\mu\nu\rho\sigma} = 0.
\label{A-eq}
\end{eqnarray}
As in $H_{\mu\nu\rho\sigma}$, if we set 
\begin{eqnarray}
F_{\mu\nu\rho\sigma}  = \theta(x) \varepsilon_{\mu\nu\rho\sigma},
\label{F}
\end{eqnarray}
with $\theta(x)$ being a scalar function, Eq. (\ref{A-eq}) requires $\theta(x)$ to be a constant
\begin{eqnarray}
\theta(x) = \theta,
\label{Theta}
\end{eqnarray}
where $\theta$ is a constant. Using this fact, Eq.   (\ref{Zero-Lagr 2}) is further simplified to be
\begin{eqnarray}
\frac{1}{16 \pi G_b}  \left[  R - 2 \Lambda_b (x) \right] + k_{1b} R^2 + k_{2b} R_{\mu\nu}^2 + k_{3b} R_{\mu\nu\rho\sigma}^2
+ {\cal{L}}_{mb} - \frac{1}{2} \theta^2 = 0.
\label{R-eq}
\end{eqnarray}

Next, under the condition of Eq. (\ref{eta-sol}), it is tedious but straightforward to derive the gravitational field equations
by taking the metric variation:
\begin{eqnarray}
&{}& \frac{1}{16 \pi G_b} \left(  G_{\mu\nu}  + \Lambda_b g_{\mu\nu} \right) + k_{1b} {}^{(1)} H_{\mu\nu} 
+ k_{2b} {}^{(2)} H_{\mu\nu} + k_{3b} {}^{(3)} H_{\mu\nu} - \frac{1}{2} T_{\mu\nu}
\nonumber\\
&+& \frac{1}{4} \cdot \frac{1}{4!} g_{\mu\nu} F_{\alpha\beta\gamma\delta}^2 
- \frac{1}{12} F_{\mu\alpha\beta\gamma} F_\nu \,^{\alpha\beta\gamma}
= 0,
\label{Eins-eq 1}
\end{eqnarray}
where $G_{\mu\nu} = R_{\mu\nu} - \frac{1}{2} g_{\mu\nu} R$ is the well-known Einstein tensor and
the energy-momentum tensor is defined by $T_{\mu\nu} = - \frac{2}{\sqrt{-g}} \frac{\delta (\sqrt{-g}  {\cal{L}}_{mb})}
{\delta g^{\mu\nu}}$ as usual.  
Furthermore, the quantities ${}^{(i)} H_{\mu\nu} (i = 1, 2, 3)$ are defined in general $n$ space-time dimensions as
\begin{eqnarray}
{}^{(1)} H_{\mu\nu} &=&  \frac{1}{\sqrt{-g}} \frac{\delta}{\delta g^{\mu\nu}} \int d^n x \sqrt{-g} R^2
\nonumber\\
&=& -2 \nabla_\mu \nabla_\nu R + 2 g_{\mu\nu} \Box R - \frac{1}{2} g_{\mu\nu} R^2 + 2 R R_{\mu\nu},
\nonumber\\
{}^{(2)} H_{\mu\nu} &=&  \frac{1}{\sqrt{-g}} \frac{\delta}{\delta g^{\mu\nu}} \int d^n x \sqrt{-g} R_{\alpha\beta}^2
\nonumber\\
&=& -2 \nabla_\rho \nabla_\mu R_\nu \, ^\rho + \Box R_{\mu\nu} + \frac{1}{2} g_{\mu\nu} \Box R 
+ 2 R_{\mu\rho} R_\nu \, ^\rho - \frac{1}{2} g_{\mu\nu} R_{\alpha\beta}^2,
\nonumber\\
{}^{(3)} H_{\mu\nu} &=&  \frac{1}{\sqrt{-g}} \frac{\delta}{\delta g^{\mu\nu}} \int d^n x \sqrt{-g} 
R_{\alpha\beta\gamma\delta}^2
\nonumber\\
&=& - \frac{1}{2} g_{\mu\nu} R_{\alpha\beta\gamma\delta}^2 
+ 2 R_{\mu\alpha\beta\gamma} R_\nu \, ^{\alpha\beta\gamma} + 4 \Box R_{\mu\nu} 
- 2 \nabla_\mu \nabla_\nu R - 4 R_{\mu\alpha} R^\alpha \, _\nu  
\nonumber\\
&+& 4 R^{\alpha\beta} R_{\alpha\mu\beta\nu},
\label{Def H}
\end{eqnarray}
where $\Box$ is a covariant d'Alembertian operator $\Box = g^{\mu\nu} \nabla_\mu \nabla_\nu$.
Note that in four space-time dimensions, 
\begin{eqnarray}
\int d^4 x \sqrt{-g} E 
\equiv \int d^4 x \sqrt{-g} ( R_{\alpha\beta\gamma\delta}^2 - 4 R_{\alpha\beta}^2 + R^2 ),
\label{Euler}
\end{eqnarray}
is a topological invariant called the Euler number, from which we have an identitiy
\begin{eqnarray}
{}^{(3)} H_{\mu\nu}  = - {}^{(1)} H_{\mu\nu} + 4 {}^{(2)} H_{\mu\nu},  
\label{H-identity}
\end{eqnarray}
so we could further simplify Eq.  (\ref{Eins-eq 1}) but for generality we will keep ${}^{(3)} H_{\mu\nu}$
in this paper though we sometimes use the fact $\delta_\mu^\mu = 4$. 

Then, using Eqs.  (\ref{F}) and  (\ref{Theta}), Eq.  (\ref{Eins-eq 1}) can be cast to the form   
\begin{eqnarray}
\frac{1}{16 \pi G_b} \left(  G_{\mu\nu}  + \Lambda_b g_{\mu\nu} \right) 
+ k_{1b} {}^{(1)} H_{\mu\nu} + k_{2b} {}^{(2)} H_{\mu\nu} + k_{3b} {}^{(3)} H_{\mu\nu}
- \frac{1}{2} T_{\mu\nu} + \frac{1}{4} \theta^2 g_{\mu\nu} 
= 0.
\label{Eins-eq}
\end{eqnarray}
Next, taking the trace of Eq.  (\ref{Eins-eq})  and using the Bianchi identity $\nabla^\mu R_{\mu\nu} 
= \frac{1}{2} \nabla_\nu R$, we have
\begin{eqnarray}
\frac{1}{16 \pi G_b} ( - R + 4 \Lambda_b ) + 2 ( 3 k_{1b} + k_{2b} + k_{3b} ) \Box R 
- \frac{1}{2} T + \theta^2 = 0,
\label{Trace-Eins-eq}
\end{eqnarray}
where $T = g^{\mu\nu} T_{\mu\nu}$.  Using Eqs. (\ref{R-eq}) and (\ref{Trace-Eins-eq}) for eliminating the term involving 
the mere scalar curvature $R$, the scalar field $c(x)$ is expressed as 
\begin{eqnarray}
c(x) &=& {\cal{L}}_{mb} - \frac{1}{2} T + \frac{\Lambda_b}{8 \pi G_b} + \frac{1}{2} \theta^2
+ k_{1b} R^2 + k_{2b} R_{\mu\nu}^2 + k_{3b} R_{\mu\nu\rho\sigma}^2
\nonumber\\
&+& 2 ( 3 k_{1b} + k_{2b} + k_{3b} ) \Box R.
\label{c(x)}
\end{eqnarray}

Thus far, our theory is completely based on the local field theories, so it is difficult to solve the cosmological constant problem
according to Weinberg \cite{Weinberg}. To tackle this hard problem, one way out is to introduce some sort of nonlocal effects
to our theory. A recipe for getting such nonlocal effects is to take the space-time average of the gravitational field equations in order to
extract information on the bare cosmological constant $\Lambda_b$. \footnote{The idea of taking the space-time average has
been also used in \cite{Linde, Tseytlin, Nima}.}  For a generic space-time dependent quantity $Q(x)$, the space-time average 
is defined as \footnote{In our previous article \cite{Oda0}, we have used the notation $\langle Q \rangle$ to describe the
space-time average. In this article, this notation is used in describing the vacuum expectation value as seen shortly.}
\begin{eqnarray}
\overline{Q(x)}  = \frac{\int d^4 x \sqrt{-g} \, Q(x)}{\int d^4 x \sqrt{-g}},
\label{ST average}
\end{eqnarray}
where the denominator $V \equiv \int d^4 x \sqrt{-g}$ denotes the space-time volume. 
With this definition, the space-time average of $c(x)$ is vanishing up to a surface term since 
\begin{eqnarray}
\overline{c(x)}  = \frac{1}{V} \int d^4 x \sqrt{-g} \, c(x) = - \frac{1}{V} \cdot \frac{1}{3!}
\int d^4 x \ \mathring{\varepsilon}^{\mu\nu\rho\sigma} \partial_{[\mu}
B_{\nu\rho\sigma]}
= 0,
\label{c-average}
\end{eqnarray}
where the definitions Eq. (\ref{4-forms}) and Eq. (\ref{H}) were utilized.
Then, taking the space-time average of Eq. (\ref{c(x)}) yields the expression for the bare cosmological constant
\begin{eqnarray}
\Lambda_b = 8 \pi G_b \left( - \overline{{\cal{L}}_{mb}} + \frac{1}{2} \overline{T} - \frac{1}{2} \theta^2
- k_{1b} \overline{R^2} - k_{2b} \overline{R_{\mu\nu}^2} - k_{3b} \overline{R_{\mu\nu\rho\sigma}^2} \right),
\label{Bare CC0}
\end{eqnarray}
where we have neglected surface terms.  Similarly, taking the space-time average of Eq.  (\ref{R-eq}), we have
\begin{eqnarray}
\Lambda_b = \frac{1}{2} \overline{R}  + 8 \pi G_b \left(  \overline{{\cal{L}}_{mb}} - \frac{1}{2} \theta^2
+ k_{1b} \overline{R^2} + k_{2b} \overline{R_{\mu\nu}^2} + k_{3b} \overline{R_{\mu\nu\rho\sigma}^2} \right),
\label{Bare CC1}
\end{eqnarray}
where we have neglected surface terms again.

In order to eliminate the bare cosmological constant $\Lambda_b$ in
the gravitational field equations (\ref{Eins-eq}), let us insert Eq. (\ref{Bare CC1}) to (\ref{Eins-eq}) whose
result reads
\begin{eqnarray}
&{}& G_{\mu\nu} + 16 \pi G_b \left( k_{1b} {}^{(1)} H_{\mu\nu} + k_{2b} {}^{(2)} H_{\mu\nu} 
+ k_{3b} {}^{(3)} H_{\mu\nu} \right)
\nonumber\\
&+& \left[ \frac{1}{2} \overline{R}  + 8 \pi G_b \left(  \overline{{\cal{L}}_{mb}}
+ k_{1b} \overline{R^2} + k_{2b} \overline{R_{\alpha\beta}^2} + k_{3b} \overline{R_{\alpha\beta\gamma\delta}^2} \right)
\right] g_{\mu\nu}
\nonumber\\
&=& 8 \pi G_b T_{\mu\nu}.
\label{Eins-eq 2}
\end{eqnarray}
Now, from Eq. (\ref{Eins-eq 2}), one can easily read off the effective cosmological constant
\begin{eqnarray}
\Lambda_{eff} = \frac{1}{2} \overline{R}  + 8 \pi G_b \left(  \overline{{\cal{L}}_{mb}}
+ k_{1b} \overline{R^2} + k_{2b} \overline{R_{\alpha\beta}^2} + k_{3b} \overline{R_{\alpha\beta\gamma\delta}^2} \right).
\label{Effective CC}
\end{eqnarray}
Next, taking the trace of Eq. (\ref{Eins-eq 2}) and then the space-time average, we obtain 
\begin{eqnarray}
\overline{R} = - 32 \pi G_b \left(  \overline{{\cal{L}}_{mb}}
+ k_{1b} \overline{R^2} + k_{2b} \overline{R_{\alpha\beta}^2} + k_{3b} \overline{R_{\alpha\beta\gamma\delta}^2} \right)
+ 8 \pi G_b \overline{T}.
\label{Average R}
\end{eqnarray}
By substituting this equation into Eq. (\ref{Effective CC}), the  effective cosmological constant can be expressed in terms of 
the space-time average over $T$
\begin{eqnarray}
\Lambda_{eff} = 8 \pi G_b \left( \frac{1}{2} \overline{T}  -  \overline{{\cal{L}}_{mb}}
- k_{1b} \overline{R^2} - k_{2b} \overline{R_{\alpha\beta}^2} - k_{3b} \overline{R_{\alpha\beta\gamma\delta}^2} \right).
\label{Effective CC-2}
\end{eqnarray}
Furthermore, using Eq. (\ref{Bare CC0}),  this effective cosmological constant can be also rewritten as 
\begin{eqnarray}
\Lambda_{eff} = \Lambda_b + 4 \pi G_b \theta^2,
\label{Effective CC-3}
\end{eqnarray}
whose expression exactly coincides with the effective cosmological constant obtained in Carroll and Remmen \cite{Carroll, Oda0}.
In this way, we have succeeded in getting three equivalent expressions for the effective cosmological constant, those are,
Eqs. (\ref{Effective CC}), (\ref{Effective CC-2}) and (\ref{Effective CC-3}).

\section{Radiative corrections to the effective cosmological constant}

The analysis done so far is purely classical, so in this section we wish to take into consideration quantum effects, in particular, radiative
corrections to the effective cosmological constant from matter fields.
Let us note that Eq. (\ref{Effective CC-2}) gives us a useful clue for evaluating the value of the cosmological constant. This effective 
cosmological constant is truly a cosmological constant controlling the Friedmann equations and other gravitational solutions, and vanishes 
for vacuum configurations of matter fields up to the higher-derivative curvature terms which also vanish for a flat Minkowski space-time. 
Since the flat Minkowski space-time is a consistent solution to the gravitational equations (\ref{Eins-eq 2}), our formulation evades 
the no-go theorem by Weinberg \cite{Weinberg} by introducing nonlocal effects.

One of the most important aspects in the cosmological constant problem is the stability problem against radiative corrections: Quantum
effects render the value of the cosmological constant infinite in the limit of the infinite cutoff when the cutoff method is used as the regularization
procedure. This radiative instability occurs even in the semiclassical theory where the gravitational field is treated classically whereas 
matter fields are treated quantum mechanically. 
The gravitational semiclassical approach is reminiscent of the successful semiclassical theory of electromagnetic dynamics where
the classical electromagnetic fields are coupled to the vacuum expectation value, which is essentially a quantum-mechanical quantity,
of the electromagnetic current operator.

It is therefore natural to ask ourselves how the effective cosmological constant changes when quantum effects are added to the
present formulation. For simplicity of the presentation but without losing generality, as the matter field, let us consider 
a massive real scalar field with nonminimal coupling:
\begin{eqnarray}
{\cal{L}}_{mb} = \frac{1}{2} \xi \phi^2 R - \frac{1}{2} g^{\mu\nu} \partial_\mu \phi \partial_\nu \phi
- \frac{1}{2} m^2 \phi^2,
\label{Massive scalar}
\end{eqnarray}
where $\xi$ and $m$ are some constants. In this article, we work with the semiclassical theory for which the matter field
$\phi(x)$ is treated quantum mechanically while both the gravitational field $g_{\mu\nu}(x)$ and the scalar field $\eta(x)$
are treated classically. In particular, we will consider the case where the scalar field  $\eta(x)$ takes the on-shell constant
value $\eta(x) = \eta$, which is a natural choice since $\eta(x) = \eta$ is the unique classical solution in our formulation.
In other words, we will perform the path integral over only the matter field $\phi(x)$.

With this physical setup, let us focus on the calculation of $\langle \overline{{\cal{L}}_{mb}} \rangle$ in $\langle \Lambda_{eff} \rangle$ 
which is the vacuum expectation value of the space-time average of the effective cosmological constant. In what follows, we
wish to prove a statement
\begin{eqnarray}
\langle \overline{{\cal{L}}_{mb}} \rangle = \overline{{\cal{L}}_{eff}},
\label{Theorem}
\end{eqnarray}
where $\overline{{\cal{L}}_{eff}}$ denotes the space-time average of an effective Lagrangian (without the factor $\eta$) obtained 
via path integral over the scalar field $\phi$.

To do that, recall that the nonlocal action  (\ref{Original CR Action}), which is nothing but the original Carroll-Remmen action, 
can be separated into two parts:
\begin{eqnarray}
S = \eta \int d^4 x \sqrt{- g} \  ( {\cal{L}}_{mb} + \tilde {\cal{L}} ),
\label{Separated CR Action}
\end{eqnarray}
where $\tilde {\cal{L}}$ denotes the Lagrangian (without the factor $\eta$) which does not include the matter field. For the present 
discussion, without loss of generality, we can set $\tilde {\cal{L}} = 0$, and we will define
\begin{eqnarray}
S = \eta \int d^4 x \sqrt{- g} \  {\cal{L}}_{mb} \equiv \eta S_{mb}. 
\label{Separated CR Action 2}
\end{eqnarray}
Then, we find 
\begin{eqnarray}
\langle \overline{{\cal{L}}_{mb}} \rangle = \frac{1}{V} \langle S_{mb} \rangle
\equiv \frac{1}{V} \frac{1}{Z} \int {\cal{D}} \phi \ S_{mb} \ \e^{i \eta S_{mb}},
\label{Logic 1}
\end{eqnarray}
where we have defined the partition function $Z$ by
\begin{eqnarray}
Z \equiv \int {\cal{D}} \phi \ \e^{i \eta S_{mb}} \equiv \e^{i \eta S_{eff}},
\label{Logic 2}
\end{eqnarray}
Now, operating $\frac{1}{i} \frac{\partial}{\partial \eta}$ on $Z$ produces
\begin{eqnarray}
\langle S_{mb} \rangle = S_{eff},
\label{Logic 3}
\end{eqnarray}
which leads to the desired statement
\begin{eqnarray}
\langle \overline{{\cal{L}}_{mb}} \rangle = \frac{1}{V} \langle S_{mb} \rangle
= \frac{1}{V}  S_{eff} = \overline{{\cal{L}}_{eff}},
\label{Logic 4}
\end{eqnarray}
where we have defined the effective Lagrangian ${\cal{L}}_{eff}$ by
\begin{eqnarray}
S_{eff} \equiv \int d^4 x  \sqrt{-g} \ {\cal{L}}_{eff}.
\label{Logic 5}
\end{eqnarray}
This statement says that evaluating $\langle \overline{{\cal{L}}_{mb}} \rangle$ is in essence equivalent to 
calculating the effective action which is obtained through path integral over only the matter field.  
 
Using dimensional regularization, it is straightforward to calculate the one-loop divergent terms in the effective
action whose result is given by \cite{Birrell, Parker}
\begin{eqnarray}
{\cal{L}}_{div} = - \frac{1}{(4 \pi)^{\frac{n}{2}}} \Biggl\{ \frac{1}{n-4} + \frac{1}{2} \left[ \gamma 
+ \log \left( \frac{m^2}{\mu^2} \right) \right] \Biggr\} \Biggl[ \frac{4 m^4}{n(n-2)} a_0 
- \frac{2 m^2}{n-2} a_1 + a_2 \Biggr],
\label{Div-term}
\end{eqnarray}
where $n$ is the dimensionality of space-time, $\gamma \approx 0.5772$ is the Euler-Mascheroni constant, and 
$\mu$ is an arbitrary renormalization mass scale. Moreover, $a_i ( i = 0, 1, 2 )$ are defined as
\begin{eqnarray}
a_0 &=& 1,  \quad  a_1 = (\frac{1}{6} - \xi) R,
\nonumber\\
a_2 &=& \frac{1}{180} R_{\alpha\beta\gamma\delta}^2 - \frac{1}{180} R_{\alpha\beta}^2
+ \frac{1}{6} (\frac{1}{5} - \xi) \Box R + \frac{1}{2} (\frac{1}{6} - \xi)^2 R^2.
\label{a_i}
\end{eqnarray}
It is evident that ${\cal{L}}_{div}$ consists of only local terms, so one can absorb it into the bare gravitational Lagrangian.
Then, the total gravitational Lagrangian becomes $\eta \sqrt{-g}$ multiplied by
\begin{eqnarray}
&-& \left( A + \frac{\Lambda_b}{8 \pi G_b} \right) + \left( B + \frac{1}{16 \pi G_b} \right) R
+ k_{1b} R^2 + k_{2b} R_{\alpha\beta}^2 + k_{3b} R_{\alpha\beta\gamma\delta}^2 
\nonumber\\
&-& \frac{a_2}{(4 \pi)^{\frac{n}{2}}} \Biggl\{ \frac{1}{n-4} + \frac{1}{2} \left[ \gamma 
+ \log \left( \frac{m^2}{\mu^2} \right) \right] \Biggr\},
\label{Renormalization}
\end{eqnarray}
where $A$ and $B$ are defined by
\begin{eqnarray}
A &=& \frac{4 m^4}{(4 \pi)^{\frac{n}{2}} n (n-2)} \Biggl\{ \frac{1}{n-2} + \frac{1}{2} \left[ \gamma 
+ \log \left( \frac{m^2}{\mu^2} \right) \right] \Biggr\},
\nonumber\\
B &=& \frac{2 m^2 (\frac{1}{6} - \xi)}{(4 \pi)^{\frac{n}{2}} (n-2)} \Biggl\{ \frac{1}{n-4} + \frac{1}{2} \left[ \gamma 
+ \log \left( \frac{m^2}{\mu^2} \right) \right] \Biggr\}.
\label{A&B}
\end{eqnarray}
Thus, the quantum effects coming from the matter fluctuations at the one-loop level can be renormalized by redefining 
the cosmological constant, Newton's constant and the coefficients in front of the higher-derivative curvature terms
from the bare quantities to the renormalized ones. For instance,
\begin{eqnarray}
\Lambda_r = \Lambda_b + 8 \pi G_b A,  \quad  G_r = \frac{G_b}{1 + 16 \pi G_b B}.
\label{Lambda&G}
\end{eqnarray}
Expression $A$ in Eq. (\ref{A&B}) has a double and single pole at $n=2$, corresponding to 
the quartically and quadratically divergent quantity,
respectively \cite{Peskin}.  Actually, there also exists the logarithmically divergent quantity in the cutoff regularization
since dimensional regularization and zeta function one are in a sense so powerful that they have the effect of hiding
some interesting terms \cite{Visser}.  And it turns out that expression $B$ has both the quadratically and 
logarithmically divergent quantities \cite{Visser}.

Moreover, the factor $a_2$ is of fourth order in derivatives of the metric tensor, so the logarithmically divergent terms 
including the factor $a_2$ are absorbed into the bare higher-derivative curvature terms to yield the renormalized 
coefficients $k_{1r}$, $k_{2r}$, and $k_{3r}$.  Of course, owing to the generalized Gauss-Bonnet theorem, in four dimensions, 
$k_{1r}$ and $k_{2r}$ are sufficient for the purpose of renormalization. Consequently, the radiative corrections associated with
the matter field $\phi(x)$ give us the following result to the part of the effective cosmological constant (\ref{Effective CC-2})
\begin{eqnarray}
- 8 \pi G_b \langle \overline{{\cal{L}}_{mb}} + k_{1b} \overline{R^2} + k_{2b} \overline{R_{\alpha\beta}^2} 
+ k_{3b} \overline{R_{\alpha\beta\gamma\delta}^2} \rangle 
= - 8 \pi G_r \left( k_{1r} \overline{R^2} + k_{2r} \overline{R_{\alpha\beta}^2} 
+ k_{3r} \overline{R_{\alpha\beta\gamma\delta}^2} \right),
\label{Quant-Effective CC-1}
\end{eqnarray}
in addition to the replacement $\Lambda_b \rightarrow \Lambda_r$. 

Note that the right hand side of Eq. (\ref{Quant-Effective CC-1}) is written in terms of finite renormalized quantities 
so that it is completely finite.  Following the same line of argument, it is easy to verify that the similar renormalization procedure
continues to the higher-loop levels. However, in the case at hand, this finiteness does not imply the quantum radiative stability 
but is just the opposite: For instance, once the cosmological constant is fixed to be a certain value at the one-loop level, 
the two-loop effect changes its finite value by an infinite value, so we need the fine-tuning at the two-loop level again, 
thereby implying the radiative instability of the effective cosmological constant under the quantum effects of matter fields.  
Thus, we conclude that the effective cosmological constant in both the nonlocal approach by Carroll and Remmen \cite{Carroll}, 
and its manifestly local approach \cite{Oda0} is unstable against radiative corrections. This fact has been also recently pointed out 
in Ref. \cite{Kaloper4} by using a different method. Incidentally, the remaining part of $\langle \Lambda_{eff} \rangle$ 
in Eq. (\ref{Effective CC-2}), which is related to the trace of the energy-momentum tensor, will be discussed in the
next section.

\section{A new geometrical model}

It is now easy to understand the reason why the effective cosmological constant is not stable under radiative corrections 
of matter fields in the Carroll-Remmen approach.  Look at the expression  (\ref{Effective CC-2}) of the effective 
cosmological constant where the matter constribution $- 8 \pi G_b \overline{{\cal{L}}_{mb}}$ is included.
This matter contribution gives rise to divergent terms in the effective action, by which we need to fine-tune various parameters
in the action at every stage in the perturbation theory, consequently causing the instability of the cosmological constant. 
Also notice that the origin of the matter contribution is traced back to the constraint  (\ref{Zero-Lagr}) which 
explicitly includes the matter Lagrangian. Thus, one plausible idea for evading the radiative instability is to 
contruct a nonlocal theory with the geometrical constraint in such a way that the effective cosmological constant does not 
include the matter contribution. 

As a concrete realization of our ideas, let us first start by presenting a manifestly local formulation of a new, geometrical
nonlocal approach.  We will neglect the higher-derivative curvature terms in this section since it turns out that they are irrelevant 
for later discussion.  A manifestly local and generally coordinate invariant action for the new nonlocal approach takes the form
\begin{eqnarray}
S = S_{New} + S_{Top},
\label{Modified Action}
\end{eqnarray}
where $S_{Top}$ is given by Eq.  (\ref{Top Action}), but the Carroll-Remmen action (\ref{CR Action}) is replaced 
with the new action $S_{New}$:
\begin{eqnarray}
S_{New}  = \int d^4 x \sqrt{- g} \ \Biggl[ \eta(x) ( R - 2 \Lambda_b ) + {\cal{L}}_{mb} 
- \frac{1}{2} \cdot \frac{1}{4!} F_{\mu\nu\rho\sigma}^2 
+ \frac{1}{6} \nabla_\mu ( F^{\mu\nu\rho\sigma} A_{\nu\rho\sigma} )
\Biggr].
\label{New Action}
\end{eqnarray}
One characteristic feature of this new action is that only the scalar curvature and the bare cosmological constant are multipled by 
the scalar field $\eta(x)$  while the total Lagrangian is so in the Carroll-Remmen action (\ref{CR Action}). 
The equations of motion for the 3-form $B_{\mu\nu\rho}$ again read Eq. (\ref{eta-eq}) and its classical solution is given 
by Eq. (\ref{eta-sol}), so we reach the nonlocal action for this new nonlocal approach
\begin{eqnarray}
S  = \int d^4 x \sqrt{- g} \ \Biggl[ \eta ( R - 2 \Lambda_b )
+ {\cal{L}}_{mb} - \frac{1}{2} \cdot \frac{1}{4!} F_{\mu\nu\rho\sigma}^2 
+ \frac{1}{6} \nabla_\mu ( F^{\mu\nu\rho\sigma} A_{\nu\rho\sigma} ) \Biggr].
\label{New nonlocal action}
\end{eqnarray}

Next, let us derive the equation of motion for the scalar field $\eta(x)$ which is this time given by
\begin{eqnarray}
\sqrt{- g} ( R - 2 \Lambda_b ) + \frac{1}{4!}  \ \mathring{\varepsilon}^{\mu\nu\rho\sigma} H_{\mu\nu\rho\sigma} = 0.
\label{New constraint}
\end{eqnarray}
Again, substituting  (\ref{H}) into the second term in this equation, we find
\begin{eqnarray}
c(x) = R - 2 \Lambda_b.
\label{c-relation}
\end{eqnarray}
Taking the space-time average of Eq.  (\ref{c-relation}) leads to
\begin{eqnarray}
\overline{R} = 2 \Lambda_b,
\label{R=2 Lambda}
\end{eqnarray}
where we have used $\overline{c(x)} =0$.  This equation might suggest that the bare cosmological constant
$\Lambda_b$ would be very tiny since we could expect $\overline{R}$ to be very small.

The equations of motion for the 3-form $A_{\mu\nu\rho}$ again yield the classical solution
Eqs. (\ref{F}) and (\ref{Theta}). Using the classical solution Eq. (\ref{eta-sol}) and (\ref{Theta}),
the gravitational equations can be derived to be
\begin{eqnarray}
\eta ( G_{\mu\nu} + \Lambda_b g_{\mu\nu} ) - \frac{1}{2} T_{\mu\nu} + \frac{1}{4} \theta^2 g_{\mu\nu} = 0.
\label{New Eins-eq 1}
\end{eqnarray}
Then, by taking the trace of Eq. (\ref{New Eins-eq 1}), one obtains
\begin{eqnarray}
\eta ( - R + 4 \Lambda_b ) - \frac{1}{2} T + \theta^2 = 0.
\label{New bare CC1}
\end{eqnarray}
Furthermore, the space-time average of this equation produces 
\begin{eqnarray}
\theta^2 = \frac{1}{2} \overline{T} - 2 \eta \Lambda_b,
\label{New bare CC2}
\end{eqnarray}
where the new constraint (\ref{R=2 Lambda}) is used.  

To eliminate $\theta^2$ in the gravitational equations, let us insert Eq.  (\ref{New bare CC2}) to (\ref{New Eins-eq 1}) 
whose result reads
\begin{eqnarray}
G_{\mu\nu}  + \left( \frac{1}{2} \Lambda_b + \frac{1}{8 \eta}  \overline{T} \right) g_{\mu\nu} = \frac{1}{2 \eta} T_{\mu\nu},
\label{New Eins-eq 2}
\end{eqnarray}
from which one can read off the effective cosmological constant
\begin{eqnarray}
\Lambda_{eff} = \frac{1}{2} \Lambda_b + \frac{1}{8 \eta} \overline{T}.
\label{New Eff CC1}
\end{eqnarray}
Using Eq.  (\ref{New bare CC2}), this effective cosmological constant can be also expressed like 
\begin{eqnarray}
\Lambda_{eff} = \Lambda_b + \frac{1}{4 \eta} \theta^2.
\label{New Eff CC2}
\end{eqnarray}
It is worthwhile to point out that the geometrical constraint (\ref{R=2 Lambda}) has produced a new effective cosmological
constant (\ref{New Eff CC1}) which does not involve the matter contribution responsible for the radiative instability. 

Now we move on to the calculation of the vacuum expectation value of the  effective cosmological
constant (\ref{New Eff CC1}). Taking the vacuum expectation value of 
(\ref{New Eff CC1}) gives us
\begin{eqnarray}
\langle \Lambda_{eff} \rangle = \frac{1}{2} \Lambda_b + \frac{1}{8 \eta} \langle \overline{T} \rangle.
\label{Quant-New Effective CC-1}
\end{eqnarray}
Note that $\eta$ is the constant mode whose origin lies in the topological term (\ref{Top Action}), so we expect that
this mode does not receive any radiative corrections. Of course, to relate the present model to a gravitational theory,
$\eta$ is needed to take the value of $\frac{1}{16 \pi G_N}$ where $G_N$ is the Newton's constant to be
determined by experiments. 
The evaluation of $\langle \overline{T} \rangle$ in the RHS needs careful considerations. 
In order to evaluate $\langle \overline{T} \rangle$ without ambiguity, as the matter field instead of a massive real scalar 
field in the Lagrangian (\ref{Massive scalar}), we need to consider free, massless generic fields conformally coupled to 
the external gravitational field $g_{\mu\nu}$ and the constant scalar field $\eta$. \footnote{At the one-loop level,
we need only free Lagrangian since the one-loop calculation amounts to evaluating the functional determinant.} 
Actually, for instance, in case of 
the non-conformally coupled, massless real scalar field, i.e., $\xi \neq \frac{1}{6}$ and $m = 0$ in (\ref{Massive scalar}),  
there appears the non-anomalous contribution \cite{Birrell}
\begin{eqnarray}
6 (\xi - \frac{1}{6}) \left[ - g^{\mu\nu} \langle \partial_\mu \phi \partial_\nu \phi \rangle
+ \xi R \langle \phi^2 \rangle \right],
\label{Non-anomaly}
\end{eqnarray}
which explicitly depends on the quantum state chosen. 

However, is it realistic to confine ourselves to only free, massless generic fields conformally coupled to the external gravitational 
field $g_{\mu\nu}$ as the matter fields?  
Then, the question which we should ask is whether matter fields in the Standard Model of elementary particles are scale-invariant? 
Relevant to this question, let us recall that it is only the (tachyonic) mass term for the Higgs field that breaks the conformal invariance 
in the Standard Model. Thus, in the absence of the Higgs mass term, the Standard Model is completely scale-invariant. 
Actually, in recent years we are witnesses of a great interest in such a "conformal" Standard Model, which belongs to physics
of beyond the Standard Model (BSM), in order to understand the Higgs mechanism more closely \cite{Bardeen, Meissner, Iso, Oda4}.  
In this BSM, the bare action is conformally invariant and the Higgs potential is generated via the Coleman-Weinberg mechanism \cite{Coleman}. 
Hence, from this new viewpoint, it is physically reasonable to work with only free, massless generic fields conformally coupled to 
the external gravitational field as the matter fields. Of course, we should refer to one important remark: In the "conformal" Standard Model,
it is anticipated that the scale symmetry is broken at the TeV scale via the Coleman-Weinberg mechanism and then the terms
which are not invariant under the scale transformation, are naturally generated, by which we have the nonzero trace part of the energy-momentum
tensor. Thus, in calculating $\langle \overline{T} \rangle$ below the TeV scale there would exist an ambiguity associated with 
the non-anomalous contributions as in Eq. (\ref{Non-anomaly}). It is an important future problem to check if the non-anomalous contributions 
give rise to some divergences or not. 

Under this situation, it is well known that at the quantum level the trace of the energy-momentum tensor has conformal
anomaly
\begin{eqnarray}
\langle T \rangle = c \, C_{\mu\nu\rho\sigma}^2 - a \, E,
\label{Conf-anomaly}
\end{eqnarray}
where $C_{\mu\nu\rho\sigma}$ is the Weyl (or conformal) tensor and $E$ is defined in Eq.  (\ref{Euler}), and
the coefficients $a$ and $c$ are given by
\begin{eqnarray}
a &=& \frac{1}{360 (4 \pi)^2} ( n_S + 11 n_D + 62 n_M ),    \nonumber\\
c &=& \frac{1}{120 (4 \pi)^2} ( n_S + 6 n_D + 12 n_M ),
\label{a-c}
\end{eqnarray}
where $n_S$, $n_D$ and $n_M$ are respectively numbers of real scalars, Dirac spinors and Maxwell fields. Note that this expression of 
conformal anomaly is exact at the one-loop level and receives no corrections from the higher-loops owing to the Adler-Bardeen theorem
\cite{Adler1, Adler2}. Then, $\langle \Lambda_{eff} \rangle$ has a rather interesting expression
\begin{eqnarray}
\langle \Lambda_{eff} \rangle &=& \frac{1}{2} \Lambda_b + \frac{1}{8 \eta} \left(  c \, \overline{C_{\mu\nu\rho\sigma}^2} 
- a \, \overline{E} \right)
\nonumber\\
&=& \frac{1}{2} \Lambda_b + \frac{1}{8 \eta} \frac{1}{V} \left(  c \int d^4 x \sqrt{-g} C_{\mu\nu\rho\sigma}^2 
- a \int d^4 x \sqrt{-g} E \right).
\label{Quant-Effective CC-3}
\end{eqnarray}
Eq. (\ref{Quant-Effective CC-3}) clearly shows that the vacuum expectation value of the effective cosmological constant 
is purely determined in terms of the geometry and topology of our universe, and is therefore stable against radiative corrections 
of the matter fields at any orders of matter loop in perturbation theory. This fact is also consistent with the expression of 
Eq. (\ref{New Eff CC2}).  

Finally, it is of interest to conjecture some things: It is nowadays well known that for most of time our universe can be 
described in terms of the spatially flat Friedmann-Lemaitre-Robertson-Walker (FLRW) cosmological model, which is an example 
of conformally flat space-times, for which the Weyl tensor identically vanishes. In addition to it, if we assume the Weyl curvature 
hypothesis by Penrose \cite{Penrose}, which insists that the Weyl tensor should vanish at the initial big bang singularity, the term
having the Weyl tensor in the RHS of Eq. (\ref{Quant-Effective CC-3}) would be zero during the almost whole history of our universe. 
About the Euler term in Eq. (\ref{Quant-Effective CC-3}), if we assume that our universe were topologically trivial,  the Euler number 
would be vanishing as well. As a result, in this case, $\langle \Lambda_{eff} \rangle = \frac{1}{2} \Lambda_b$, which means 
the vanishing of radiative corrections to the cosmological constant. On the other hand, if topology of our universe were nontrivial, 
the Euler number would play a role in the quantum effects to the cosmological constant. This phenomenon reminds us of the
wormhole scenario by Coleman \cite{Coleman2}.

\section{Discussions}

In this article, we have studied quantum aspects of the nonlocal approach to the cosmological constant \cite{Carroll} to some degree on the
basis of its manifestly local and generally coordinate invariant formulation which has been recently developed by the present author \cite{Oda0}. 
In particular, we have investigated radiative corrections to the effective cosmological constant from matter fields which are coupled to 
external graviational field and the scalar field. We have found that the quantum effects change the Newton's constant, the cosmological 
constant, the coefficients of the higher-derivative curvature terms from the bare quantities to the renormalized ones via renormalization. 
Consequently, we have verified that  in the original Carroll-Remmen nonlocal approach, the effective cosmological constant, 
which is truly the cosmological constant controlling the Friedmann equation etc., is unstable against radiative corrections.

Motivated with this radiative instability of the Carroll-Remmen nonlocal approach, we have proposed a new geometrical approach
where the new geometrical constraint does not include the matter contribution at all.  
It has been explicitly shown that the vacuum expectation value of the effective cosmological constant has a contribution from
the conformal anomaly and its value is completely determined by the geometry and topology of our universe.

It is well known that in four dimensions the conformal anomaly consists of two types of anomaly term, one of which is of topological
character and is proportional to the Euler density whose integral over four dimensional manifolds is called the Euler number. 
This $a$-conformal anomaly plays a critical role in the $a$-theorem \cite{Cardy, Komargodski}, which is a four-dimensional analog 
of the two-dimensional $c$-theorem \cite{Zamolodchikov}.  Since our expression of the effective cosmological constant naturally involves 
a space-time integral of the trace of the energy-momentum tensor, this term becomes precisely the Euler number distinguishing
the topology of the universe and suggests a similar dynamical mechanism to the Coleman's wormhole scenario \cite{Coleman2}. 

On the other hand, the other type of anomaly term is proportional to the square of the Weyl curvature. This $c$-conformal anomaly
is not related to the Cardy's $a$-theorem but might be related to the Penrose's Weyl curvature hypothesis \cite{Penrose}, by which 
the initial big bang is constrained in a way that the resulting universe model closely resembles the FLRW model in its early stages and 
there would be the second law of thermodynamics in the universe that we observe.  It is of interest that in our model 
both the Coleman's wormhole scenario \cite{Coleman2} and the Penrose's Weyl curvature hypothesis \cite{Penrose} appear 
in a natural way, which might support the physical plausibility of our approach to the cosmological constant problem.

An important future work is to include quantum effects stemming from graviton loops in the present analysis. We wish to return this 
problem in near future.

\begin{flushleft}
{\bf Acknowledgements}
\end{flushleft}
This work is supported in part by the Grant-in-Aid for Scientific 
Research (C) No. 16K05327 from the Japan Ministry of Education, Culture, 
Sports, Science and Technology.


\end{document}